\documentstyle[amsfonts,aps]{revtex}
\begin{document}
\title{Decoherence and the Final Pointer Basis.}
\author{Mario Castagnino, Roberto Laura}
\address{Instituto de F\'{\i}sica de Rosario, CONICET-UNR\\
Av. Pellegrini 250, 2000 Rosario, Argentina.\\
E-mail: laura@ifir.ifir.edu.ar}
\maketitle

\begin{abstract}
Using a functional method it is demonstrated that a generic quantum system
evolves to a decohered state in a final pointer basis.
\end{abstract}

\section{Introduction.}

We will demonstrate that for a wide set of quantum systems the quantum
regime can be consider as the transient phase while the final classical
equilibrium regime is the permanent state. We will find a basis where exact
matrix decoherence appears for these final states. Therefore we will find a
set of final intrinsically consistent histories.

\section{Decoherence.}

\subsection{Decoherence in the energy.}

Let us consider a closed and isolated quantum system with $N+1$ dynamical
variables and a Hamiltonian endowed with a continuous spectrum and just one
bounded ground state. So the discrete part of the spectrum of $H$ has only
one value $\omega _0$ and the continuous spectrum is let say $0\leq \omega
<\infty $. Eventually we will give the collective name $x$ to both $\omega
_0 $ and $\omega .$ Let us assume that it is possible to diagonalize the
Hamiltonian $H$, together with $N$ observables $O_i$ ($i=1,...,N)$. The
operators ($H$, $O_1$,...,$O_N$) form a {\it complete set of commuting
observables} (CSCO). For simplicity we also assume a discrete spectrum for
the $N$ observables $O_i$. Therefore we write 
\begin{equation}
H=\omega _0\sum_m|\omega _0,m\rangle \langle \omega _0,m|+\int_0^\infty
\omega \sum_m|\omega ,m\rangle \langle \omega ,m|d\omega  \label{2.4}
\end{equation}
where $\omega _0<0$ is the energy of the ground state, and $m\doteq
\{m_1,...,m_N\}$ labels a set of discrete indexes which are the eigenvalues
of the observables $O_1$,...,$O_N$. $\{|\omega _0,m\rangle ,|\omega
,m\rangle \}$ is a basis of simultaneous generalized eigenvectors of the
CSCO: 
\begin{eqnarray*}
H|\omega _0,m\rangle &=&\omega _0|\omega _0,m\rangle ,\quad H|\omega
,m\rangle =\omega |\omega ,m\rangle , \\
O_i|\omega _0,m\rangle &=&m_i|\omega _0,m\rangle ,\quad O_i|\omega ,m\rangle
=m_i|\omega ,m\rangle .
\end{eqnarray*}

The most general observable that we are going to consider in our model
reads: 
\begin{eqnarray}
O &=&\sum_{mm^{\prime }}O(\omega _0)_{mm^{\prime }}|\omega _0,m\rangle
\langle \omega _0,m^{\prime }|+\sum_{mm^{\prime }}\int_0^\infty d\omega
O(\omega )_{mm^{\prime }}|\omega ,m\rangle \langle \omega ,m^{\prime }|+ 
\nonumber \\
&&+\sum_{mm^{\prime }}\int_0^\infty d\omega O(\omega ,\omega _0)_{mm^{\prime
}}|\omega ,m\rangle \langle \omega _0,m^{\prime }|+  \nonumber \\
&&+\sum_{mm^{\prime }}\int_0^\infty d\omega ^{\prime }O(\omega _0,\omega
^{\prime })_{mm^{\prime }}|\omega _0,m\rangle \langle \omega ^{\prime
},m^{\prime }|+  \nonumber \\
&&+\sum_{mm^{\prime }}\int_0^\infty \int_0^\infty d\omega d\omega ^{\prime
}O(\omega ,\omega ^{\prime })_{mm^{\prime }}|\omega ,m\rangle \langle \omega
^{\prime },m^{\prime }|,  \label{2.5}
\end{eqnarray}
where $O(\omega )_{mm^{\prime }}$, $O(\omega ,\omega _0)_{mm^{\prime }}$, $%
O(\omega _0,\omega )_{mm^{\prime }}$ and $O(\omega ,\omega ^{\prime
})_{mm^{\prime }}$ are ordinary functions of the real variables $\omega $
and $\omega ^{\prime }$(these functions must have some mathematical
properties in order to develop the theory; these properties are listed in
paper \cite{CyLII}). We will say that these observables belong to a space $%
{\cal O}$. This space has the {\it basis} $\{|\omega _0,mm^{\prime })$, $%
|\omega ,mm^{\prime })$, $|\omega \omega _0,mm^{\prime })$, $|\omega
_0\omega ^{\prime },mm^{\prime })$, $|\omega \omega ^{\prime },mm^{\prime
})\}$: 
\[
|\omega _0,mm^{\prime })\doteq |\omega _0,m\rangle \langle \omega
_0,m^{\prime }|,\quad |\omega ,mm^{\prime })\doteq |\omega ,m\rangle \langle
\omega ,m^{\prime }|, 
\]
\begin{equation}
|\omega \omega _0,mm^{\prime })\doteq |\omega ,m\rangle \langle \omega
_0,m^{\prime }|,\quad |\omega _0\omega ^{\prime },mm^{\prime })\doteq
|\omega _0,m\rangle \langle \omega ^{\prime },m^{\prime }|,  \label{2.5'}
\end{equation}
\[
|\omega \omega ^{\prime },mm^{\prime })\doteq |\omega ,m\rangle \langle
\omega ^{\prime },m^{\prime }| 
\]
The quantum states $\rho $ are measured by the observables just defined,
computing the mean values of these observable in the quantum states, i. e.
in the usual notation: $\langle O\rangle _\rho =Tr(\rho ^{\dagger }O)$ \cite
{Ballentine}. These mean values, generalized as in paper \cite{CyLII}, can
be considered as linear functionals $\rho $ (mapping the vectors $O$ on the
real numbers)$,$ that we can call $(\rho |O)$ \cite{Bogo}. In fact, this is
a generalization of the usual mean value definition. Then $\rho \in {\cal %
S\subset O}^{^{\prime }},$ where ${\cal S}$ is a convenient convex set
contained in ${\cal O}^{^{\prime }}$, the space of linear functionals over $%
{\cal O}$ \cite{CyLI}, \cite{CyLIII}. The basis of ${\cal O}^{\prime }$
(that can also be considered as the {\it co-basis} of ${\cal O)}$ is $%
\{(\omega _0,mm^{\prime }|$, $(\omega ,mm^{\prime }|$, $(\omega \omega
_0,mm^{\prime }|$, $(\omega _0\omega ^{\prime },mm^{\prime }|$, $(\omega
\omega ^{\prime },mm^{\prime }|\}$ defined as functionals by the equations: 
\[
(\omega _0,mm^{\prime }|\omega _0,nn^{\prime })=\delta _{mn}\delta
_{m^{\prime }n^{\prime }},\quad (\omega ,mm^{\prime }|\eta ,nn^{\prime
})=\delta (\omega -\eta )\delta _{mn}\delta _{m^{\prime }n^{\prime }}, 
\]
\[
(\omega \omega _0,mm^{\prime }|\eta \omega _0,nn^{\prime })=\delta (\omega
-\eta )\delta _{mn}\delta _{m^{\prime }n^{\prime }}, 
\]
\[
(\omega _0\omega ^{\prime },mm^{\prime }|\omega _0\eta ^{\prime },nn^{\prime
})=\delta (\omega ^{\prime }-\eta ^{\prime })\delta _{mn}\delta _{m^{\prime
}n^{\prime }}, 
\]
\begin{equation}
(\omega \omega ^{\prime },mm^{\prime }|\eta \eta ^{\prime },nn^{\prime
})=\delta (\omega -\eta )\delta (\omega ^{\prime }-\eta ^{\prime })\delta
_{mn}\delta _{m^{\prime }n^{\prime }}.  \label{2.5''}
\end{equation}
and all other $(.|.)$ are zero. Then, a generic quantum state reads: 
\begin{eqnarray}
\rho &=&\sum_{mm^{\prime }}\overline{\rho (\omega _0)}_{mm^{\prime }}(\omega
_0,mm^{\prime }|+\sum_{mm^{\prime }}\int_0^\infty d\omega \overline{\rho
(\omega )}_{mm^{\prime }}(\omega ,mm^{\prime }|+  \nonumber \\
&&+\sum_{mm^{\prime }}\int_0^\infty d\omega \overline{\rho (\omega ,\omega
_0)}_{mm^{\prime }}(\omega \omega _0,mm^{\prime }|+  \nonumber \\
&&+\sum_{mm^{\prime }}\int_0^\infty d\omega ^{\prime }\overline{\rho (\omega
_0,\omega ^{\prime })}_{mm^{\prime }}(\omega _0\omega ^{\prime },mm^{\prime
}|+  \nonumber \\
&&+\sum_{mm^{\prime }}\int_0^\infty d\omega \int_0^\infty d\omega ^{\prime }%
\overline{\rho (\omega ,\omega ^{\prime })}_{mm^{\prime }}(\omega \omega
^{\prime },mm^{\prime }|  \label{2.6}
\end{eqnarray}
where 
\[
\overline{\rho (\omega ,\omega _0)}_{mm^{\prime }}=\rho (\omega _0,\omega
)_{m^{\prime }m},\quad \overline{\rho (\omega ,\omega ^{\prime })}%
_{mm^{\prime }}=\rho (\omega ^{\prime },\omega )_{m^{\prime }m}, 
\]
and $\overline{\rho (\omega _0)}_{mm}$ and $\overline{\rho (\omega )}_{mm}$
are real and non negative satisfying the total probability condition 
\begin{equation}
(\rho |I)=\sum_m\rho (\omega _0)_{mm}+\sum_m\int_0^\infty d\omega \rho
(\omega )_{mm}=1,  \label{2.6'}
\end{equation}
where $I=\sum_m|\omega _0,m\rangle \langle \omega _0,m|+\int_0^\infty
d\omega \sum_m|\omega ,m\rangle \langle \omega ,m|$ is the identity operator
in ${\cal O}$. Eq. (\ref{2.6'}) is the extension to state functionals of the
usual condition $Tr\rho ^{\dagger }=1$, used when $\rho $ is a density
operator.

The time evolution of the quantum state $\rho $ reads: 
\begin{eqnarray}
\rho (t) &=&\sum_{mm^{\prime }}\overline{\rho (\omega _0)}_{mm^{\prime
}}(\omega _0,mm^{\prime }|+\sum_{mm^{\prime }}\int_0^\infty d\omega 
\overline{\rho (\omega )}_{mm^{\prime }}(\omega ,mm^{\prime }|+  \nonumber \\
&&+\sum_{mm^{\prime }}\int_0^\infty d\omega \overline{\rho (\omega ,\omega
_0)}_{mm^{\prime }}e^{i(\omega -\omega _0)t}(\omega \omega _0,mm^{\prime }|+
\nonumber \\
&&+\sum_{mm^{\prime }}\int_0^\infty d\omega ^{\prime }\overline{\rho (\omega
_0,\omega ^{\prime })}_{mm^{\prime }}e^{i(\omega _0-\omega ^{\prime
})t}(\omega _0\omega ^{\prime },mm^{\prime }|+  \nonumber \\
&&+\sum_{mm^{\prime }}\int_0^\infty d\omega \int_0^\infty d\omega ^{\prime }%
\overline{\rho (\omega ,\omega ^{\prime })}_{mm^{\prime }}e^{i(\omega
-\omega ^{\prime })t}(\omega \omega ^{\prime },mm^{\prime }|  \label{2.7}
\end{eqnarray}

As we only measure mean values of observables in quantum states, i. e.: 
\begin{eqnarray}
\langle O\rangle _{\rho (t)} &=&(\rho (t)|O)=  \nonumber \\
&=&\sum_{mm^{\prime }}\overline{\rho (\omega _0)}_{mm^{\prime }}O(\omega
_0)_{mm^{\prime }}+\sum_{mm^{\prime }}\int_0^\infty d\omega \overline{\rho
(\omega )}_{mm^{\prime }}O(\omega )_{mm^{\prime }}+  \nonumber \\
&&+\sum_{mm^{\prime }}\int_0^\infty d\omega \overline{\rho (\omega ,\omega
_0)}_{mm^{\prime }}e^{i(\omega -\omega _0)t}O(\omega ,\omega _0)_{mm^{\prime
}}+  \nonumber \\
&&+\sum_{mm^{\prime }}\int_0^\infty d\omega ^{\prime }\overline{\rho (\omega
_0,\omega ^{\prime })}_{mm^{\prime }}e^{i(\omega _0-\omega ^{\prime
})t}O(\omega _0,\omega ^{\prime })_{mm^{\prime }}+  \nonumber \\
&&+\sum_{mm^{\prime }}\int_0^\infty d\omega \int_0^\infty d\omega ^{\prime }%
\overline{\rho (\omega ,\omega ^{\prime })}_{mm^{\prime }}e^{i(\omega
-\omega ^{\prime })t}O(\omega ,\omega ^{\prime })_{mm^{\prime }},
\label{2.8}
\end{eqnarray}
using the Riemann-Lebesgue theorem we obtain the limit, for all $O\in {\cal O%
}$ 
\begin{equation}
\lim_{t\rightarrow \infty }\langle O\rangle _{\rho (t)}=\langle O\rangle
_{\rho _{*}}  \label{2.9}
\end{equation}
where we have introduced the diagonal asymptotic or equilibrium state
functional 
\begin{equation}
\rho _{*}=\sum_{mm^{\prime }}\overline{\rho (\omega _0)}_{mm^{\prime
}}(\omega _0,mm^{\prime }|+\sum_{mm^{\prime }}\int_0^\infty d\omega 
\overline{\rho (\omega )}_{mm^{\prime }}(\omega ,mm^{\prime }|  \label{2.10}
\end{equation}
Therefore, in a weak sense we have: 
\begin{equation}
W\lim_{t\rightarrow \infty }\rho (t)=\rho _{*}  \label{2.11}
\end{equation}
Thus, any quantum state goes weakly to a linear combination of the energy
diagonal states $(\omega _0,mm^{\prime }|$ and $(\omega ,mm^{\prime }|$ (the
energy off-diagonal states $(\omega \omega _0,mm^{\prime }|$, $(\omega
_0\omega ^{\prime },mm^{\prime }|$ and $(\omega \omega ^{\prime },mm^{\prime
}|$ are not present in $\rho _{*}$). This is the case if we observe and
measure the system evolution with {\it any possible observable of space }$%
{\cal O}${\it .} Then, from the observational point of view, we have
decoherence of the energy levels, even that, from the strong limit point of
view the off-diagonal terms never vanish, they just oscillate, since we
cannot directly use the Riemann-Lebesgue theorem in the operator equation (%
\ref{2.7}).

\subsection{Decoherence in the other ''momentum'' dynamical variables.}

Having established the decoherence in the energy levels we must consider the
decoherence in the other dynamical variables $O_i$, of the CSCO where we are
working. We will call these variables ''momentum variables''. For the sake
of simplicity we will consider, as in the previous section, that the spectra
of these dynamical variables are discrete. As the expression of $\rho _{*}$
given in eq. (\ref{2.10}) involve only the time independent components of $%
\rho (t)$, it is impossible that a different decoherence process take place
to eliminate the off-diagonal terms in the remaining $N$ dynamical
variables. Therefore, the only thing to do is to find if there is a basis
where the off-diagonal components of $\rho (\omega _0)_{mm^{\prime }}$ and $%
\rho (\omega )_{mm^{\prime }}$ vanish at any time before the equilibrium is
reached.

Let us consider the following change of basis 
\[
|\omega _0,r\rangle =\sum_mU(\omega _0)_{mr}|\omega _0,m\rangle ,\qquad
|\omega ,r\rangle =\sum_mU(\omega )_{mr}|\omega ,m\rangle , 
\]
where $r$ and $m$ are short notations for $r\doteq \{r_1,...,r_N\}$ and $%
m\doteq \{m_1,...,m_N\}$, and $\left[ U(x)^{-1}\right] _{mr}=\overline{U(x)}%
_{rm}$ ($x$ denotes either $\omega _0<0$ or $\omega \in {\Bbb R}^{+}$).

The new basis $\{|\omega _0,r\rangle ,|\omega ,r\rangle \}$ verify the
generalized orthogonality conditions 
\begin{eqnarray*}
\langle \omega _0,r|\omega _0,r^{\prime }\rangle &=&\delta _{rr^{\prime
}},\quad \langle \omega ,r|\omega ^{\prime },r^{\prime }\rangle =\delta
(\omega -\omega ^{\prime })\delta _{rr^{\prime }}, \\
\langle \omega _0,r|\omega ,r^{\prime }\rangle &=&\langle \omega ,r|\omega
_0,r^{\prime }\rangle =0.
\end{eqnarray*}

As $\overline{\rho (\omega _0)}_{mm^{\prime }}=\rho (\omega _0)_{m^{\prime
}m}$ and $\overline{\rho (\omega )}_{mm^{\prime }}=\rho (\omega )_{m^{\prime
}m}$, it is possible to choose $U(\omega _0)$ and $U(\omega )$ in such a way
that the off-diagonal parts of $\rho (\omega _0)_{rr^{\prime }}$ and $\rho
(\omega )_{rr^{\prime }}$ vanish, i.e. 
\[
\rho (\omega _0)_{rr^{\prime }}=\rho _r(\omega _0)\,\delta _{rr^{\prime
}},\qquad \rho (\omega )_{rr^{\prime }}=\rho _r(\omega )\,\delta
_{rr^{\prime }}. 
\]
Therefore, there is a{\it \ final pointer basis} for the observables given
by $\{|\omega _0,rr^{\prime })$, $|\omega ,rr^{\prime })$, $|\omega \omega
_0,rr^{\prime })$, $|\omega _0\omega ^{\prime },rr^{\prime })$, $|\omega
\omega ^{\prime },rr^{\prime })\}$ and defined as in eq. (\ref{2.5'}). The
corresponding final pointer basis for the states $\{(\omega _0,rr^{\prime }|$%
, $(\omega ,rr^{\prime }|$, $(\omega \omega _0,rr^{\prime }| $, $(\omega
_0\omega ^{\prime },rr^{\prime }|$, $(\omega \omega ^{\prime },rr^{\prime
}|\}$ diagonalizes the time independent part of $\rho (t)$ and therefore it
diagonalizes the final state $\rho _{*}$%
\begin{equation}
\rho _{*}=W\lim_{t\rightarrow \infty }\rho (t)=\sum_r\rho _r(\omega
_0)(\omega _0,rr|+\sum_r\int_0^\infty d\omega \rho _r(\omega )(\omega ,rr|.
\label{RO1}
\end{equation}

Now we can define the final {\it exact pointer observables} \cite{Zurek} 
\begin{equation}
P_i=\sum_rP_r^i(\omega _0)|\omega _0,r\rangle \langle \omega
_0,r|+\int_0^\infty d\omega \sum_rP_r^i(\omega )|\omega ,r\rangle \langle
\omega ,r|.  \label{RO2}
\end{equation}
As $H$ and $P_i$ are diagonal in the basis $\{|\omega _0,r\rangle $, $%
|\omega ,r\rangle \}$, the set $\{H,P_i,...P_N\}$ is precisely the complete
set of commuting observables (CSCO) related to this basis, where $\rho _{*}$
is diagonal in the corresponding co-basis for states. For simplicity we
define the operators $P_i$ such that $P_r^i(\omega _0)=P_r^i(\omega )=r_i$,
thus 
\begin{equation}
P_i|\omega _0,r\rangle =r_i|\omega _0,r\rangle ,\qquad P_i|\omega ,r\rangle
=r_i|\omega ,r\rangle .  \label{RO3}
\end{equation}
Therefore $\{|\omega _0,r\rangle $, $|\omega ,r\rangle \}$ is the observers
pointer basis where there is a perfect decoherence in the corresponding
state co-basis. Moreover the generalized states $(\omega _0,rr|$ and $%
(\omega ,rr|$ are constants of the motion, and therefore these exact pointer
observables have a constant statistical entropy and will be ''at the top of
the list'' of Zurek's ''predictability sieve'' \cite{Zurek}.

Therefore:

i.- Decoherence in the energy is produced by the time evolution.

ii.- Decoherence in the other dynamical variables can be seen if we choose
an adequate basis, namely the final pointer basis.

Our main result is eq. (\ref{RO1}): {\it When }$t\rightarrow \infty $ {\it %
then }$\rho (t)\rightarrow \rho _{*}$ {\it and in this state the dynamical
variables }$H,P_1,...,P_N$ {\it are well defined. Therefore the eventual
conjugated variables to these momentum variables (namely: configuration
variables, if they exists) are completely undefined.}

In fact, calling by ${\Bbb L}_i$ the generator of the displacements along
the eventual configuration variable conjugated to $P_i$, we have $({\Bbb L}%
_i\rho _{*}|O)=(\rho _{*}|{\Bbb L}_i^{\dagger }O)=(\rho _{*}|[P_i,O])=0$ for
all $O\in {\cal O}$. Then $\rho _{*}$\ is homogeneous in these configuration
variables.

From the preceding section we may have the feeling that the process of
decoherence must be found in all the physical systems, and therefore, all of
them eventually would become classical when $\hbar \rightarrow 0$. It is not
so as explained in \cite{Deco}.

\section{The classical equilibrium limit.}

\subsection{Expansion in sets of classical motions.}

In this section we will use the Wigner integrals that introduce an
isomorphism between quantum observables $O$ and states $\rho $ and their
classical analogues $O^W(q,p)$ and $\rho ^W(q,p)$ \cite{Wigner}: 
\begin{eqnarray}
O^W(q,p) &=&\int d\lambda \,\langle q-\frac \lambda 2|O|q+\frac \lambda 2%
\rangle \,\exp (\frac{i\lambda p}\hbar )  \nonumber \\
\rho ^W(q,p) &=&\frac 1{\pi \hbar }\int d\lambda \,(\rho ||q+\lambda \rangle
\langle q-\lambda |)\,\exp (\frac{2i\lambda p}\hbar ).  \label{Wig}
\end{eqnarray}

It is possible to prove that $\int dq\,dp\,\rho ^W(q,p)=(\rho |I)=1$, but $%
\rho ^W$ is not in general non negative. It is also possible to deduce that 
\begin{equation}
(\rho ^W|O^W)=\int dq\,dp\,\rho ^W(q,p)O^W(q,p)=(\rho |O),  \label{mean}
\end{equation}
and therefore to the mean value in the classical Liouville space it
corresponds the mean value in the quantum Liouville space.

Moreover, calling by $L$ the classical Liouville operator, and by ${\Bbb L}$
the quantum Liouville-Von Neumann operator, we have 
\begin{equation}
L\left[ \rho ^W(q,p)\right] =\left[ {\Bbb L}\rho \right] ^W(q,p)+O(\hbar ),
\label{ara1}
\end{equation}
where $L\,\rho ^W(q,p)=i\left\{ H^W(q,p),\rho ^W(q,p)\right\} _{PB}$ and 
\begin{equation}
({\Bbb L}\rho |O)=(\rho |[H,O]).  \label{ara2}
\end{equation}
Finally, if $O=O_1O_2$, where $O_1$ and $O_2$ are two quantum observables,
we have 
\begin{equation}
O^W(q,p)=O_1^W(q,p)O_2^W(q,p)+O(\hbar ).  \label{pro}
\end{equation}

We will prove that the distribution function $\rho _{*}^W(q,p)$, that
corresponds to the state functional $\rho _{*}$ via the Wigner integral is a
non negative function of the classical constants of the motion, in our case $%
H^W(q,p)$, $P_1^W(q,p)$,..., $P_N^W(q,p),$ obtained from the corresponding
quantum operators $H$, $P_1$,..., $P_N$.

From eq. (\ref{RO1}) we have: 
\begin{equation}
\rho _{*}=W\lim_{t\rightarrow \infty }\rho (t)=\sum_r\rho _r(\omega
_0)(\omega _0,rr|+\sum_r\int_0^\infty d\omega \rho _r(\omega )(\omega ,rr|,
\label{5.2}
\end{equation}
so we must compute: 
\begin{equation}
\rho _{\omega r}^W(q,p)\doteq \pi ^{-1}\int (\omega ,rr||q+\lambda \rangle
\langle q-\lambda |)e^{2ip\lambda }d\lambda  \label{5.3}
\end{equation}
We know from \cite{CyLII} section II. C, (or we can prove directly from eqs.(%
\ref{RO1}-\ref{RO3})) that 
\begin{eqnarray}
(\omega _0,rr|H^n) &=&\omega _0^n,\quad (\omega ,rr|H^n)=\omega ^n, 
\nonumber \\
(\omega _0,rr|P_i^n) &=&r_i^n,\quad (\omega ,rr|P_i^n)=r_i^n,\quad i=1,...,N
\label{5.4}
\end{eqnarray}

for $n=0,1,2,...$ Using the relation (\ref{pro}) between quantum and
classical products of observables and relation (\ref{mean}) between quantum
and classical mean values, in the limit $\hbar \rightarrow 0$ (we will
consider that we always take this limit when we refer to classical equations
below) we deduce that the characteristic property of the distribution $\rho
_{\omega r}^W(q,p)$, that corresponds to the state functional $(\omega ,rr|$%
, is: 
\begin{equation}
\int \rho _{\omega r}^W(q,p)[H^W(q,p)]^ndqdp=\omega ^n,\quad \int \rho
_{\omega r}^W(q,p)[P_i^W(q,p)]^ndqdp=r_i^n,  \label{5.5}
\end{equation}
for any natural number $n.$ Thus $\rho _{\omega r}^W(q,p)$ must be the
functional 
\begin{equation}
\rho _{\omega r}^W(q,p)=\delta (H^W(q,p)-\omega )\delta
(P_1^W(q,p)-r_1)...\delta (P_N^W(q,p)-r_N).  \label{5.6}
\end{equation}
For the distribution $\rho _{\omega _0r}^W(q,p)$ corresponding to the state
functional $(\omega _0,rr|$, we obtain 
\begin{equation}
\rho _{\omega _0r}^W(q,p)=\delta (H^W(q,p)-\omega _0)\delta
(P_1^W(q,p)-r_1)...\delta (P_N^W(q,p)-r_N).  \label{5.6'}
\end{equation}
Therefore, going back to eq. (\ref{5.2}) and since the Wigner relation is
linear, we have: 
\begin{equation}
\rho _{*}^W(q,p)=\sum_r\rho _r(\omega _0)\rho _{\omega
_0r}^W(q,p)+\sum_r\int_0^\infty d\omega \rho _r(\omega )\rho _{\omega
r}^W(q,p).  \label{5.7}
\end{equation}
Also we obtain $\rho _{*}^W(q,p)\geq 0$, because $\rho _r(\omega _0)$ and $%
\rho _r(\omega )$ are non negative.

Therefore, the classical state $\rho _{*}^W(q,p)$ is a linear combination of
the generalized classical states $\rho _{xr}^W(q,p)$ (where $x$ is either $%
\omega _0$ or $\omega $), having well defined values $x$, $r_1$,..., $r_N$
of the classical observables $H^W(q,p)$, $P_1^W(q,p)$,..., $P_N^W(q,p)$ and
the corresponding classical canonically conjugated variables completely
undefined since $\rho _{xr}^W(q,p)$ is not a function of these variables. 
{\it So we reach, in the classical case, to the same conclusion than in the
quantum case (see end of subsection 2. 2). }But now all the classical
canonically conjugated variables $a_0,a_1,...,a_N$ do exist since they can
be found solving the corresponding Poisson brackets differential equations.
We can also expand the densities given in eqs. (\ref{5.6}-\ref{5.7}) in
terms of classical motions as shown in \cite{Deco}.

\section{Conclusion.}

i.- We have shown that the quantum state functional $\rho (t)$ evolves to a
diagonal state $\rho _{*}$.

ii.- This quantum state $\rho _{*}$ has $\rho _{*}^W(q,p)$ as its
corresponding classical density.

iii.- This classical density can be decomposed in sets of classical motions
where $H^W$, $P_1^W$,..., $P_N^W$ remain constant. These motions have
origins $a_0(0),a_1(0),...,a_N(0)$ distributed in an homogeneous way.

iv.- From eqs. (\ref{5.6}-\ref{5.7}) we obtained that 
\[
\rho _{*}^W(q,p)=f(H^W(q,p),P_1^W(q,p),...,P_N^W(q,p))\geq 0. 
\]


\begin{references}
\bibitem{CyLII}  Laura R., Castagnino M.{\it ,} Phys. Rev. A, {\bf 57},
4140-4152,1998.

\bibitem{Ballentine}  Ballentine L. E.,{\it \ Quantum mechanics,} Prentice
Hall, Englewoods Cliffs, 1990.

\bibitem{Bogo}  Bogolubov N. N., Logunov A. A., Todorov I. J., {\it %
Introduction to axiomatic quantum field theory,} Benjamin, London, 1975.

\bibitem{CyLI}  Castagnino M., Laura R., Phys. Rev. A, {\bf 56,} 108-119,
1997.

\bibitem{CyLIII}  Castagnino M., Gunzig E., {\it A minimal irreversible
quantum mechanic: the axiomatic formalism, }Int. Journ. Theo. Phys, in
press, 1999.

\bibitem{GyHUCS}  Gell-Mann M., Hartle J. USCSRTH-94-09, 1994.

\bibitem{Deco}  Castagnino M., Laura R, {\it Functional Approach to Quantum
Decoherence and the Classical Equilibrium Limit, }Submitted to Phys. Rev.
D., 1999.

\bibitem{Wigner}  Hillery M., O'Connell R. F., Scully M. O., Wigner E. P.,
Phys. Rep., {\bf 106,} 123, 1984.

\bibitem{Zurek}  Zurek W. H., {\it Preferred sets of states, predictability,
classicality, and environment- induced decoherence. In ''Physical Origin of
Time Asymmetry}'', Halliwell J. J. et al. eds., Cambridge University Press,
Cambridge, 1994.
\end{references}
\end{document}